\documentclass[a4paper]{article}
\usepackage[english]{babel}
\usepackage[utf8]{inputenc}

\usepackage{fancybox}
\usepackage{pdfpages}
\usepackage{stmaryrd}
\usepackage{amssymb}
\usepackage{amsmath}
\usepackage{amsthm}
\usepackage{mathabx}
\usepackage{todonotes}
\usepackage{tikz}
\usetikzlibrary{shapes,arrows}

\usepackage{alltt}
\usepackage{url}
\usepackage{listings}
\usepackage{enumitem}

\newtheorem{theorem}{Theorem}
\newtheorem{corollary}{Corollary}[theorem]

\begin{document}

\newcommand{\forceindent}{\leavevmode{\parindent=1em\indent}}

\def\Name{\ensuremath{\text{\bf Name}}}
\def\OR{\ensuremath{\ |\ }}
\def\TO{\ensuremath{\rightarrow}}
\def\FROM{\ensuremath{\leftarrow}}
\def\LB{\ensuremath{\llbracket}}
\def\RB{\ensuremath{\rrbracket}}

\newcommand\LIT[1]{\ensuremath{\text{\tt #1}}}
\newcommand\SLIT[1]{\ \LIT{#1}\ }
\newcommand\IF[3]{\LIT{if}\ #1\ \LIT{then}\ #2\ \LIT{else}\ #3}
\newcommand\APP[2]{#1\ #2}
\newcommand\CASE[4]{\LIT{case}\ #1:#2\ \LIT{of}\ #3 \TO #4}
\newcommand\INBR[1]{\ensuremath{\llbracket #1 \rrbracket}}
\newcommand\MGU[2]{\LIT{unify}(#1, #2)}
\newcommand\INVERT[1]{(\LIT{invert}\ #1)}

\newcommand\EvalWith[4]{\ensuremath{ \Delta#1 #2 \vdash #3 \downarrow #4}}
\newcommand\Eval[2]{\EvalWith{}{\Gamma}{#1}{#2}}
\newcommand\LaveWith[4]{\ensuremath{ \Delta#1 #2 \vdash #3 \uparrow #4}}
\newcommand\Lave[2]{\LaveWith{}{\Gamma}{#1}{#2}}

\newcommand\Because[4]{\ensuremath{\Delta{#1}\INBR{#2 \downarrow #3}\rightsquigarrow #4}}
\newcommand\Esuaceb[4]{\ensuremath{\Delta{#1}\INBR{#2 \uparrow #3}\leftsquigarrow #4}}

\newcommand\LinearHasType[4]{\ensuremath{\Delta#1 #2 \vdash #3 : #4}}
\newcommand\LinearEpytSah[4]{\ensuremath{\Delta#1 #2 \vDash #3 : #4}}
\newcommand\LinearBindTypes[4]{\ensuremath{\Delta#1 | #2 : #3 \Downarrow #4}}
\newcommand\LinearUnBindTypes[4]{\ensuremath{\Delta#1 | #2 : #3 \Uparrow #4}}

\newcommand\UPARROW[1]{$\uparrow$#1}
\newcommand\DOWNARROW[1]{$\downarrow$#1}
\newcommand\RIGHTARROW[1]{$\rightarrow$#1}
\newcommand\LEFTARROW[1]{$\leftarrow$#1}
\newcommand\TYPE[1]{$\tau$#1}
\newcommand\LinINFER[1]{$\Downarrow$#1}
\newcommand\LinUNFER[1]{$\Uparrow$#1}

\newcommand \BX [1]
  {\scriptsize\framebox{{\raisebox{0pt}[0.7\baselineskip][0.01\baselineskip]{\small #1}}}}

\newcommand\Axiom[2]
                 {\ensuremath{\text{\small #1}:\frac{\displaystyle}
                 {\displaystyle #2}}
                 }
\newcommand\InfOne[3]
                 {\ensuremath{\text{\small #2}:\frac{\displaystyle #1}
                 {\displaystyle #3}}
                 }
\newcommand\InfTwo[4]
                 {\ensuremath{\text{\small #3}:\frac{\displaystyle #1 \quad #2}
                 {\displaystyle #4}}
                 }
\newcommand\InfThree[5]
                 {\ensuremath{\text{\small #4}:
                     \frac{\displaystyle #1 \quad #2 \quad #3}
                          {\displaystyle #5}}
                 }
\newcommand\InfFour[6]
                 {\ensuremath{\text{\small #5}:
                     \frac{\displaystyle #1 \quad #2 \quad #3 \quad #4}
                          {\displaystyle #6}}
                 }
\newcommand\InfFive[7]
                 {\ensuremath{\text{\small #6}:
                     \frac{\displaystyle #1 \quad #2 \quad #3 \quad #4 \quad #5}
                          {\displaystyle #7}}
                 }


\title{Jeopardy: An Invertible Functional Programming Language}

\author{Joachim Tilsted Kristensen$^1$ \and Robin Kaarsgaard$^2$\footnote{Supported by DFF--International Postdoctoral Fellowship 0131-00025B.
}, \and Michael Kirkedal Thomsen$^{1,3}$}

\date{$^1$ University of Oslo, Norway\\
$^2$ University of Edinburgh, UK\\
$^3$ University of Copenhagen, Denmark}

\maketitle

\begin{abstract}
  Algorithms are ways of mapping problems to solutions. An algorithm
  is invertible precisely when this mapping is injective, such that the
  initial problem can be uniquely inferred from its solution.

  While invertible algorithms can be described in general-purpose
  languages, no guarantees are generally made by such languages as
  regards invertibility, so ensuring invertibility requires additional
  (and often non-trivial) proof. On the other hand, while
  \emph{reversible} programming languages guarantee that their
  programs are invertible by restricting the permissible operations to
  those which are locally invertible, writing programs in the
  reversible style can be cumbersome, and may differ significantly
  from conventional implementations even when the implemented
  algorithm is, in fact, invertible.

  In this paper we introduce Jeopardy, a functional programming
  language that guarantees program invertibility \emph{without}
  imposing local reversibility.  In particular, Jeopardy allows the
  limited use of uninvertible -- and even nondeterministic! --
  operations, provided that they are used in a way that can be
  statically determined to be invertible.  To this end, we outline an
  \emph{implicitly available arguments analysis} and three further
  approaches that can give a partial static guarantee to the
  (generally difficult) problem of guaranteeing invertibility.
\end{abstract}

\section{Introduction}
Reversible programming languages guarantee program invertibility by
enforcing a strict syntactic discipline: programs are comprised only
of parts which are themselves immediately invertible (\emph{locally
invertible}), and these parts can only be combined in ways which
preserve invertibility. In this way, the \emph{global} problem of
ensuring the invertibility of an entire program is reduced to a
\emph{local} problem of ensuring the invertibility of its parts.

However, writing algorithms in the reversible style can be cumbersome:
in a certain sense, it corresponds to requiring that programmers
provide machine checkable proofs that their algorithms are
invertible. As such, writing programs in these languages requires some
experience, and can in some cases be notoriously hard. To mitigate the
problem, this work investigates a more relaxed approach to reversible
language design that requires only (global) invertibility.

We present the language Jeopardy; a functional language bearing
syntactic resemblance to you garden variety functional programming
language and which exhibits the expected semantics for programs running in
the conventional direction. However, in order to support program
inversion for a particular class of morally reversible programs that
fail the syntactic condition of reversibility, we also seek to extend
the semantics of Jeopardy to be \emph{relational} in a conservative
way. For example, consider the following program and its (manually
implemented) inverse:

\begin{lstlisting}
 swap p =              swap-inverse (b, a) =
   let a = first  p in   let p = (invert second) b in
   let b = second p in   let p = (invert first ) a in
   (b, a).               p.
\end{lstlisting}

\noindent Not all parts of \LIT{swap} are invertible: for
instance, \LIT{first} is not invertible at all. Nonetheless,
\LIT{swap} clearly describes an invertible algorithm, as all of the
information needed to reconstruct its input is contained in its
output.

To strengthen our intuition about why \LIT{swap} is invertible, let us
inspect the possible ways of implementing \LIT{first} or \LIT{second}. Both
have to throw some information away, like so:
\begin{lstlisting}
  first  (a, _) = a.   first-inverse  a = (a, _).
  second (_, b) = b.   second-inverse b = (_, b).
\end{lstlisting}

\noindent Because of this explicit deletion of data, \LIT{first} and
\LIT{second} are not information preserving transformations and it is
this data loss that makes \LIT{first} and \LIT{second}
non-invertible. However, when considered together as above, we see
that the \emph{``open''} part of their inverse function outputs (i.e.,
the underscore $\_$, which can be thought of as a unification
variable) can always be unified with a \emph{``closed''} term (not
containing such) from the other function.

Though there is a great deal of overlap between reversibility and
invertibility, the notion of reversibility, as known from reversible
computations, is philosophically distinct from the current work. In
particular, we do not ask that programs are implemented only using small
locally invertible parts, and can as such be cleanly mapped to a reversible
low-level abstract machine or reversible hardware.

\paragraph{Related work}
Program inversion~\cite{McCarthy:GenTest,Dijkstra:ProgramInv} concerns
the automatic synthesis of the inverse to a given program (if such an
inverse exists), while inverse execution seeks to interpret inverse
programs from forward programs directly. Like compilation and
interpration, the two are connected by a \emph{Futamura
projection}~\cite{AbramovGlueck:Universal,Futamura:Projections}.
Reversible programming~\cite{YokoyamaGlueck:2007:Janus,
  YokoyamaAxelsenGlueck:2012:LNCS, JacobsenEtal:2018,
  KirkedalKaarsgaardSoeken:Ricercar, HeunenKaarsgaardKarvonen:Arrows,
  JamesSabry:2014:RC,ThomsenAxelsen:2016:IFL}, program inversion, and inverse execution have
seen applications in areas as diverse as
debugging~\cite{GiachinoLaneseMezzina:RevDeb,
  LaneseNishidaPalaciosVidal:CauDEr}, high-performance
simulation~\cite{SchordanOppelstrupThomsenGlueck:DiscEventSim,
  SchordanJeffersonBarnesOppelstrupQuinlan:DiscEventSim}, quantum
computing~\cite{HeunenKaarsgaard:QuantInfEff,
  HeunenKaarsgaard:BennettStinespring}), and
robotics~\cite{SchultzBordignonStoy:RevExSelfRec,LaursenSchultzEllekilde:AutoRecovery},
and is intimately connected reversible model of
computation~\cite{Landauer:1961,Bennett:1973,Huffman:1959}.

\paragraph{Structure}
In the following section (Section~\ref{sec:language-description}) we
introduce Jeopardy and its syntax. Section~\ref{sec:reversible-semantics}
will detail the reversible semantics, which includes rules for both forward
and backward interpretation, while Section~\ref{sec:invertible-semantics}
suggests various strategies for conservatively relaxing the reversible
semantics. Finally in Section~\ref{sec:conclusion} we conclude on what we
have learned thus far.

An implementation of Jeopardy can be found at~\cite{KristensenJeopardyRepo}.

Jeopardy is a minimalistic first order functional language with
user-definable algebraic data types and inverse function invocation.  The
latter is invoked by the special keyword \texttt{invert}.  The syntax for
algebraic datatype declaration differs slightly from the norm, in that a sum
of products has to be declared using the keyword \texttt{data} rather than
denoted directly in the program using the symbol $\cdot{}+\cdot{}$. This may
seem odd to some theoretic computer scientists, but is common notation among
programmers. The full grammar can be found in Figure~\ref{fig:syntax}.

\begin{figure}
\begin{center}
\begin{align*}
x          &\in \Name                                        &\text{(Well-formed variable names).}\\
c          &\in \Name                                        &\text{(Well-formed constructor names).}\\
\tau       &\in \Name                                        &\text{(Well-formed datatype names).}\\
f          &\in \Name                                        &\text{(Well-formed function names).}\\
p          &::= [c\ p_i] \OR x                               &\text{(Patterns).}\\
v          &::= [c\ v_i]                                     &\text{(Values).}\\
\Delta     &::= f\ (p : \tau_p) : \tau_t\ =\ t\ .\ \Delta    &\text{(Function definition).}\\
           &\OR \LIT{data}\ \tau\ =\ [c\ \tau_i]_j\ .\ \Delta &\text{(Data type definition).}\\
           &\OR \LIT{main}\ g\ .                             &\text{(Main function declaration).}\\
g          &::= f \OR \INVERT{g}                             &\text{(Function).}\\
t          &::= p                                            &\text{(Patterns in terms).}\\
           &\OR g\ p                                         &\text{(Function application).}\\
           &\OR \CASE{t}{\tau}{p_i}{t_i}                     &\text{(Case statement).}
\end{align*}
\end{center}
\caption{The syntax of Jeopardy.}
\label{fig:syntax}
\end{figure}

To clarify, a pattern $p$ is either a variable, or a constructor applied to
(possibly 0) other patterns. A value $v$ is a pattern that does not contain
any variables. A program $\Delta$, is a list of mutually recursive function
and datatype definitions, followed by a main function declaration. Functions
are described by a name $f$, an input pattern, two type annotations (one for
input and one for output), and a term $t$ describing the functions body. A
term is either a pattern, an application, or a case statement that branches
execution. Application is special, because the operator may be a function
symbol, denoting conventional application, or ``\LIT{invert}'' of a function
symbol, calling its inverse function from the corresponding inverted
program.

Running a program in the conventional direction corresponds to calling the
declared main function on a value provided by the caller in an empty
context. Similarly, running a program backwards corresponds to calling the
main function's inverse on said value. Since function application is a term,
reasoning about programs is reasoning about terms; as such, we will focus on
terms from here on out. The syntax of terms has been designed to be small in
order to make reasoning easier, at the cost of making programs harder to
read and write. In the interest of writing intuitive program examples, we
will use a couple of derived syntactic connectives that depend on the
program $\Delta$ in which they are written, shown in Figure~\ref{fig:sugar}.

\begin{figure}
\begin{align*}
  \LB [c\ t_i] \RB_{\Delta[\LIT{data}\ \tau = [c \tau_i]_j]} &:= \CASE{t_i}{\tau_i}{p_i}{[c\ p_i]}\\
  \LB (t_1, t_2) \RB_\Delta &:= \LB [\LIT{pair}\  t_1\ t_2]\RB_\Delta\\
  \LB t_1 : t_2 \RB_\Delta &:= \LB [\LIT{cons}\ t_1\ t_2]\RB_\Delta\\
  \LB \LIT{[]} \RB_\Delta &:= [\LIT{nil}]\\
  \LB \APP{f}{t}\RB_{\Delta[f (\cdot : \tau) : \cdot = \cdot]} &:= \CASE{t}{\tau}{p}{\APP{f}{p}}\\
  \LB \LIT{let}\ p : \tau =\ t\ \LIT{in}\ t' \RB_{\Delta} &:=
  \CASE{t}{\tau}{p}{t'}\\
  \LB t' \RB_{\Delta[f (p_i : \tau_p) : \tau_t = t_i.]} &:= \LB t' \RB_{\Delta[f (x : \tau_p) : \tau_t = \CASE{x}{\tau_p}{p_i}{t_i}{}]}
\end{align*}
\caption{Disambiguation of syntactic sugar.}
\label{fig:sugar}
\end{figure}

\subsection{Examples}
\label{sec:examples}

In order to motivate the need for an invertible functional programming
language, the following section compares programs written in the reversible
style, to show that Jeopardy programs can get much closer to a conventional
way of writing these programs. Suppose that we declared two datatypes
\LIT{nat} and \LIT{pair} as follows:

\begin{lstlisting}
  data nat  = [zero] [suc nat].
  data pair = [pair nat nat].
\end{lstlisting}

\noindent It is not uncommon to want to be able to add a pair of numbers, so we write
an algorithm to do just that:

\begin{lstlisting}
  add ([zero ], n) = n
  add ([suc k], n) = add (k, [suc n]).
\end{lstlisting}

\noindent
Our algorithm \LIT{add} is not invertible, because it does not describe a
bijective function. To be precise, \LIT{add} is not injective, since
\begin{align*}
\text{\LIT{add ([suc [zero]], [zero])} = \LIT{add ([zero], [suc [zero]])}}
\end{align*}
\noindent
even though
\begin{align*}
\text{\LIT{([suc [zero]], [zero])}} \neq \text{\LIT{([zero], [suc [zero]])}}
\end{align*}
\noindent
However, since \LIT{add} \textit{is} linearly typeable, the
corresponding RFun and CoreFun programs will evaluate to a runtime error (as
they should) upon calling \LIT{add} with an \LIT{m} that differs from
\LIT{[zero]}, as the result will not be syntactically orthogonal to the
variable pattern \LIT{n}.

Now, suppose the caller of \LIT{add} happens to know the value of \LIT{m} in
advance; meaning that \LIT{m} will be available in the future from the
perspective of the inverse program. Then the case to use for inverse
interpretation suddenly become unambiguous: either \LIT{m} is \LIT{[zero]}
and we can unambiguously choose the first case, or \LIT{m} is \LIT{[suc k]},
the caller now knows \LIT{k}, and can deterministically uncall \LIT{add}
recursively in the second case. Therefore, the corresponding Jeopardy
program will allow the programmer to call \LIT{add} in a context that knows
about \LIT{m} as exemplified below:

\begin{lstlisting}
  fibber (m, n) = (add (m, n), m).

  fib-pair [zero ] = ([suc [zero]], [suc [zero]]).
  fib-pair [suc k] = fibber (fib-pair k).

  fib n = (n, (first (fib-pair n))).

  main fib.
\end{lstlisting}

\noindent
Here, the function \LIT{fib} computes the pair containing $n$ and the $n$'th
Fibonacci number. Notice that it will never be possible to write an
invertible implementation of the Fibonacci function that does not include
something extra in its output, since the first couple of outputs has to be
\LIT{[suc [zero]]} for two different inputs. The helper function
\LIT{fib-pair} becomes deterministic by the same ``\textit{trick}'' as
explained for \LIT{add}. Additionally, notice that even though the \LIT{n}
is part of the output of \LIT{fib}, it is insufficient to uncompute
\LIT{fib} by projecting out the input argument, since we agreed already,
that \LIT{first} is not invertible. So, either we have to check that the
second part of the output indeed does compute from \LIT{n} in the
conventional direction, or we need to infer a unique environment in which
the output was computed as we will discuss in Section~\ref{sec:reversible-semantics}.

Before moving on to doing so, we will further our intuition about what can
be decided about invertible programs by considering the example of
implementing map which has been specialised\footnote{The problem of
  extending an invertible (or even a reversible) programming language with
  (real) higher order functions will be worthy of its own paper.} to apply a
specific function \texttt{f}:

\begin{lstlisting}
  data list = [nil] [cons nat list].
  data pair = [pair list list].

  reverse ([]    , ys) = ys.
  reverse (y : xs, ys) = reverse (xs, y : ys).

  f ... = ...

  map-f-iter ([]    , ys) = reverse (ys, []).
  map-f-iter (x : xs, ys) = map-f (xs, f x : ys).

  map-f xs = map-f-iter (xs, []).

  main map-f.
\end{lstlisting}

\noindent
This definition of \texttt{map-f} is clearly invertible: As
\texttt{map-f-iter} and \texttt{reverse} both move elements from the first
to the second component of their input, as such, the first component always
becomes smaller, so both algorithms terminate in their respective first
cases. Moreover their second component is always initially
empty. Consequently, uncalling \texttt{map-f} will have to uncall
\texttt{map-f-iter} (and hence also \texttt{reverse}) with an empty list as
the second component.

However, for an interpreter to make this conclusion requires a non-trivial
program analysis which cannot necessarily be performed inductively over the
syntax of the program in general. In this particular case, the core
difference between \texttt{fibber} and \texttt{map-f}, is that the context
in which \texttt{fibber} calls \texttt{add}, provides information about the
variable that \texttt{add} is branching on, making reverse pattern matching
deterministic. \texttt{map-f-iter} on the other hand, is branching on
\texttt{xs}, about which \texttt{map-f} does not syntactically provide any
information about. However, the cases do become syntactically orthogonal
when elaborating on the variable that \texttt{map-f} \textit{does} provide
information about:

\begin{lstlisting}
map-f-iter ([]    ,     ys) = reverse (ys, []).
map-f-iter (x : xs, y : ys) = map-f (xs, f x : y : ys).
map-f-iter (x : xs,     []) = map-f (xs, f x :     []).
\end{lstlisting}

\noindent
Additionally, unrolling \texttt{map-f-iter} yielded a particularly
interesting result, because it suggests two unique entry and exit
conditions for a Janus style loop:

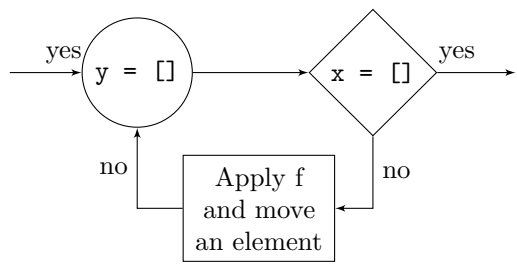
\begin{center}
\tikzstyle{assert} = [draw, circle, node distance=3cm, minimum height=4em]
\tikzstyle{condition} = [diamond, draw, text width=4em, text badly centered, node distance=3cm, inner sep=0pt, minimum height=4em]
\tikzstyle{statement} = [rectangle, draw, text width=5em, text centered, minimum height=4em]
\tikzstyle{line} = [draw, -latex']

\begin{tikzpicture}[node distance = 1cm, auto]
    \node [] (start) {};
    \node [assert, right of=start,node distance=1.8cm] (as) {\texttt{y = []}};
    \node [right of=as,node distance=1.6cm] (mid) {};
    \node [condition, right of=mid,node distance=1.5cm] (cond) {\texttt{x = []}};
    \node [right of=cond,node distance=2cm] (final) {};
    \node [statement, below of=mid,node distance=1.8cm] (stmt) {Apply f and move an element};

    \path [line] (start) -- node [near end] {yes} (as) ;
    \path [line] (as) --  (cond) ;
    \path [line] (cond) |- node [near start] {no} (stmt) ;
    \path [line] (cond) -- node [near start] {yes} (final) ;
    \path [line] (stmt) -| node [near end] {no} (as) ;
\end{tikzpicture}

\end{center}

Suggesting a way of obtaining tail-recursion optimisation for functional
reversible programs that look like tail recursive functions in a
conventional programming style. Just like reversible higher-order functions,
this has been left as future work.

\section{Reversible semantics}
\label{sec:reversible-semantics}

Aside from the special keyword \LIT{invert}, which invokes inverse
interpretation of functions, Jeopardy is purposefully limited to garden
variety syntactic constructs. It is desired that these constructs mean the
conventional thing when interpreted in the conventional direction, and that
more exotic constructions such as RFun's \LIT{rlet}-statement, can be
derived by combining \LIT{invert} with the case-statement and function
application. It is likewise desired that such constructions mean the
expected thing when inversely interpreted. In this section we present a
reversible operational semantics for Jeopardy, in which all function
definitions must be locally invertible.

The goal of this exercise is to ensure that invertible algorithms written in
the reversible style, are invertible in the same sense as that of
corresponding programs, written in languages that force programmers to
formulate their algorithms in this way -- In Section~\ref{sec:invertible-semantics}, we proceed to explaining how the reversible
semantics can be relaxed to only require global invertibility. That is, to
require a programs main function to be invertible, but not necessarily any
other functions.

This reversible semantics is inspired by those of RFun and CoreFun, and
the main difference is a separation of concerns. For instance, the judgement
rules of RFun can be read in two different ways. First, when read in the
conventional direction, a term is evaluated to a value in an environment.
Second, in the other direction, a resulting value is used to search a term for
the unique environment in which that term would have yielded that particular
result.  In this regard, the semantics of Jeopardy programs are
operationalised by four mutually recursive judgements;
Figures~\ref{fig:evaluation} and~\ref{fig:inverse-evaluation} show the
judgements for interpretation, with and against the conventional direction,
while figures~\ref{fig:inference} and \ref{fig:inverse-inference} describe
an algorithm for inferring unique environments under which linear terms (and
inverted linear terms) evaluated to their canonical forms.  Likewise, type
checking has been factored out into four mutually recursive
judgements. Figures~\ref{fig:linear-typing} and~\ref{fig:inverse-linear-typing} about linear typing (and inverse linear
typing), and Figures~\ref{fig:variable-type-bindings} and~\ref{fig:inverse-variable-type-bindings} about inferring the unique
environment in which a linear term is typeable.

The motivation for producing explicit operational semantics for
interpretation in both directions, is to enable more fine grained program
analysis, such as the ones outlined in Section~\ref{sec:invertible-semantics}.  The remainder of this section will cover
the meaning of each judgement of the reversible semantics, and towards the
end of this section, we provide the meta theoretic properties that this
semantics guarantee local invertibility on functions. To start somewhere,
programs conventionally run from ``\textit{top to bottom}''. The
corresponding judgement rules are therefore denoted by a downwards pointing
arrow, and the form is \Eval{t}{v}, where $\Delta$ is a copy of the program
text, $\Gamma$ is a mapping between variable names and values, and the
judgement reads ``\textit{in $\Delta$, $\Gamma$ stands witness that $t$
  evaluates to $v$ in the conventional direction}''.
\begin{enumerate}
  \item The rules \DOWNARROW{Variable}, \DOWNARROW{Constructor} and
    \DOWNARROW{Application} are the usual rules for looking up variables and
    applying first order functions. \texttt{unify/2} is the most general
    unifier as usual as well.
  \item\label{rule:eval-case}
    The \DOWNARROW{Cases} rule says that when the selector term $t$
    evaluates to a value $v_i$, and $v_i$ unifies with the $i$th pattern
    $p_i$, and the $i$th term $t_i$ evaluates to some value $v$ under the
    bindings from the unification, then the whole term evaluates to $v$. The
    side condition $\psi$ is an abbreviation for the bidirectional
    first-match policy, namely that we require $p_j$ not to unify with $v_i$
    whenever $j < i$ holds, and that \Because{}{t_j}{v_i}{\Gamma} should not
    hold for all such $j$ either.
  \item The \DOWNARROW{Inversion} rule invokes inverse interpretation for
    function application, which can be seen in Figure~\ref{fig:inverse-evaluation}.
\end{enumerate}

\begin{figure}
  \begin{flushleft}
    \fbox{\Eval{t}{v}} (for $t$ closed under $\Gamma$)
  \end{flushleft}
\begin{center}
  \Axiom
  {\DOWNARROW{Variable}}    
                                 {\Eval{x}{v}} $(\Gamma(x) = v)$
  \quad
  \InfOne                        {\Eval{p_i}{v_i}}
  {\DOWNARROW{Constructor}} 
                                 {\Eval{[c\ p_i]}{[c\ v_i]}}

  \InfTwo                        {\Eval{t}{v_i}}
                                 {\EvalWith{}{(\Gamma\circ{\MGU{v_i}{p_i}})}{t_i}{v}}
  {\DOWNARROW{Cases}}       
                                 {\Eval{\CASE{t}{\tau}{p_i}{t_i}}{v}}
                                 $(\psi)$

  \InfTwo                        {\Eval{p}{v'}}
                                 {\EvalWith{}{(\MGU{v'}{p'})}{t'}{v}}
  {\DOWNARROW{Application}} 
                                 {\EvalWith{[f\ (p' : \cdot) : \cdot = t']}{\Gamma}{\APP{f}{p}}{v}}
  \quad
  \InfOne                        {\Lave{\APP{g}{p}}{v}}
  {\DOWNARROW{Inversion}}   
                                 {\Eval{\APP{\INVERT{g}}{p}}{v}}
\end{center}
\caption{Interpretation in the conventional direction. The side condition
  ($\psi$) denotes the bidirectional first match policy as detailed in
  bullet point (\ref{rule:eval-case}).}
\label{fig:evaluation}
\end{figure}

\begin{figure}
  \begin{flushleft}
    \fbox{\Lave{t}{v}} (for $t$ closed under $\Gamma$)
  \end{flushleft}
\begin{center}
  \InfThree                      {\Eval{p}{v'}}
                                 {\Because{}{t'}{v'}{\Gamma'}}
                                 {\EvalWith{}{\Gamma'}{p'}{v}}
  {\UPARROW{Application}} 
                                 {\LaveWith{[f\ (p' : \cdot) : \cdot =\ t']}{\Gamma}{f\ p}{v}}
  \quad
  \InfOne                        {\Eval{g\ p}{v}}
  {\UPARROW{Inversion}}   
                                 {\Lave{(\LIT{invert}\ g)\ p}{v}}
\end{center}

\caption{Inverse interpretation for linear programs.}
\label{fig:inverse-evaluation}
\end{figure}

Since the rules for inverse interpretation run in the opposite direction of
the conventional one, their names have been annotated with an arrow pointing
upwards. Furthermore, since the keyword \LIT{invert} can only appear in an
application, there are only two rules.
\begin{enumerate}[resume]
\item The \UPARROW{Inversion} rule says that inverting inverse
  interpretation is to resume computation in the conventional direction; Not
  much to see here.
\item Finally, in the \UPARROW{Application} rule, the looming problem of
  inverse interpretation of functional programs emerges from hiding. The
  rule says that if the argument to an inverse function evaluated to some
  value $v'$, and the body of the corresponding function in the source
  program evaluated to $v'$ in the conventional direction because of a
  unique environment $\Gamma'$, then the result of the inverse function is
  the arguments for the function in the source program as evaluated under
  $\Gamma'$.
\end{enumerate}

\begin{figure}
  \begin{flushleft}
    \fbox{\Because{}{t}{v}{\Gamma}} (for linear terms $t$)
  \end{flushleft}
\begin{center}
  \Axiom
  {\RIGHTARROW{Variable}}     
                                 {\Because{}{x}{v}{\{x \mapsto v\}}}
  \quad
  \InfOne                        {\Because{}{p_i}{v_i}{\Gamma_i}}
  {\RIGHTARROW{Constructor}}  
                                 {\Because{}{[c\ p_i]}{[c\ v_i]}{\circ{\Gamma_i}}}

  \InfThree                      {\Because{}{t_i}{v_i}{\Gamma_i}}
                                 {\EvalWith{}{\Gamma_i}{p_i}{v}}
                                 {\Because{}{t}{v}{\Gamma}}
  {\RIGHTARROW{Cases}}        
                                 {\Because{}{\CASE{t}{\tau}{p_i}{t_i}}{v_i}{\Gamma_i\circ\Gamma}}
                                 $(\psi)$

  \InfThree                      {\Because{}{t'}{v}{\Gamma'}}
                                 {\EvalWith{}{\Gamma'}{p'}{v'}}
                                 {\Because{}{p}{v'}{\Gamma}}
  {\RIGHTARROW{Application}}  
                                 {\Because{[f\ (p' : \cdot) : \cdot =\ t']}{\APP{f}{p}}{v}{\Gamma}}
  \quad
  \InfOne                        {\Esuaceb{}{\APP{g}{p}}{v}{\Gamma}}
  {\RIGHTARROW{Inversion}}    
                                 {\Because{}{\APP{\INVERT{g}}{p}}{v}{\Gamma}}
\end{center}
\caption{Environment inference for linear programs. Again, the side
  condition ($\psi$) denotes the bidirectional first match policy as noted
  in bullet point (\ref{rule:infer-case}).}
\label{fig:inference}
\end{figure}

\begin{figure}
  \begin{flushleft}
    \fbox{\Esuaceb{}{t}{v}{\Gamma}} (for linear terms $t$)
  \end{flushleft}
\begin{center}
  \InfThree                      {\Because{}{p'}{v}{\Gamma'}}
                                 {\EvalWith{}{\Gamma'}{t'}{v'}}
                                 {\Because{}{p}{v'}{\Gamma}}
  {\LEFTARROW{Application}} 
                                 {\Esuaceb{[f\ p'\ =\ t']}{\APP{f}{p}}{v}{\Gamma}}
  \quad
  \InfOne                        {\Because{}{\APP{g}{p}}{v}{\Gamma}}
  {\LEFTARROW{Inversion}}   
                                 {\Esuaceb{}{\APP{\INVERT{g}}{p}}{v}{\Gamma}}
\end{center}
\caption{Inverse environment inference.}
\label{fig:inverse-inference}
\end{figure}

We can hide the problem of searching for $\Gamma'$ in the rules inverse
interpretation. But in the interest of separating concerns, we have separate
rules about searching for context. The form is \Because{}{t}{v}{\Gamma} and
it reads ``\textit{in $\Delta$, the linear term $t$ evaluated to $v$ because
  of the unique environment $\Gamma$}'', and the details can be found in
Figure~\ref{fig:inference}:
\begin{enumerate}[resume]
  \item The \RIGHTARROW{Variable} rule says that if $x$ was a linear term,
    and it evaluated to $v$, it must have been because of the unique
    environment, containing a single binding $x \mapsto v$.
  \item The \RIGHTARROW{Constructor} rule says that the unique environment
    under which a constructor evaluated to a value is the composition of the
    unique (and disjoint by linearity) environments under which its parts
    evaluated.
  \item\label{rule:infer-case}
    The \RIGHTARROW{Cases} rule, still requires the bidirectional first
    match policy $\psi$. See \DOWNARROW{Cases} (bullet point
    (\ref{rule:eval-case})).
  \item The \RIGHTARROW{Application} and \RIGHTARROW{Inversion} rules can be
    found in Figure~\ref{fig:inverse-inference}.
\end{enumerate}
Because the rules for environment inference require linearity, we have given
typing rules that are usual for linear typing ~\cite{Wadler:1990,
  JacobsenEtal:2018}. The main judgement form is
\LinearHasType{}{\Sigma}{t}{\tau} and it reads, ``\textit{in the program
  $\Delta$, the term $t$ has type $\tau$ under $\Sigma$}'', where $\Sigma$
is a mapping between variable names and type names. Moreover, just like the
rules for interpretation, typing has a typing environment inference
algorithm with the form \LinearBindTypes{}{t}{\tau}{\Sigma} which reads
``\textit{In the program $\Delta$, we know that the linear term $t$ has type
  $\tau$ because of the unique typing environment $\Sigma$}''.


\begin{figure}
  \begin{flushleft}
    \fbox{\LinearHasType{}{\Sigma}{t}{\tau}}
  \end{flushleft}
\begin{center}
  \Axiom
  {\TYPE{Variable}}    
                         {\LinearHasType{}{\{x \mapsto \tau\}}{x}{\tau}}

  \InfOne                {\LinearHasType{}{\Sigma_i}{p_i}{\tau_i}}
  {\TYPE{Constructor}} 
                         {\LinearHasType
                           {[\LIT{data}\ \tau = [c\ \tau_i]_j]}
                           {(\circ \Sigma_i)}
                           {[c\ p_i]}
                           {\tau}
                         }

  \InfThree              {\LinearHasType{}{\Sigma}{t}{\tau}}
                         {\LinearBindTypes{}{p_i}{[c\ \tau_i]}{\Sigma_i}}
                         {\LinearHasType{}{\Sigma_i \circ \Sigma_j}{t_i}{\tau'}}
  {\TYPE{Cases}}       
                         {\LinearHasType
                           {[\LIT{data}\ \tau = [c\ \tau_i]_j]}
                           {(\Sigma \circ \Sigma_j)}
                           {\CASE{t}{\tau}{p_i}{t_i}}
                           {\tau'}
                         }

  \InfThree              {\LinearHasType{}{\Sigma}{p'}{\tau_p}}
                         {\LinearBindTypes{}{p}{\tau_p}{\Sigma_p}}
                         {\LinearHasType{}{\Sigma_p}{t}{\tau_t}}
  {\TYPE{Application}} 
                         {\LinearHasType
                           {[f (p : \tau_p) : \tau_t = t]}
                           {\Sigma}
                           {\APP{f}{p'}}
                           {\tau_t}
                         }

  \InfOne                {\LinearEpytSah{}{\Sigma}{\APP{g}{p}}{\tau}}
  {\TYPE{Inversion}}   
                         {\LinearHasType{}{\Sigma}{\APP{\INVERT{g}}{p}}{\tau}}
\end{center}
\caption{Linear typing.}
\label{fig:linear-typing}
\end{figure}

\begin{figure}
  \begin{flushleft}
    \fbox{\LinearEpytSah{}{\Sigma}{t}{\tau}}
  \end{flushleft}
\begin{center}
  \InfThree                      {\LinearEpytSah{}{\Sigma}{p'}{\tau_t}}
                                 {\LinearUnBindTypes{}{t}{\tau_t}{\Sigma_t}}
                                 {\LinearHasType{}{\Sigma_t}{p}{\tau_p}}
  {\TYPE{InverseApplication}}     
                                 {\LinearEpytSah{[f\ (p : \tau_p) : \tau_t = t]}{\Sigma}{\APP{f}{p'}}{\tau_p}}

  \InfOne                        {\LinearHasType{}{\Sigma}{\APP{g}{p}}{\tau}}
  {\TYPE{InverseInversion}} 
                                 {\LinearEpytSah{}{\Sigma}{\APP{\INVERT{g}}{p}}{\tau}}
\end{center}
\caption{Inverse linear typing.}
\label{fig:inverse-linear-typing}
\end{figure}


\begin{figure}
  \begin{flushleft}
    \fbox{\LinearBindTypes{}{p}{\tau}{\Sigma}}
  \end{flushleft}
\begin{center}
  \Axiom
  {\LinINFER{Variable}}  
                           {\LinearBindTypes{}{x}{\tau}{\{x \mapsto \tau\}}}

  \InfOne                  {\LinearBindTypes{}{p_i}{\tau_i}{\Sigma_i}}
  {\LinINFER{Constructor}}   
                           {\LinearBindTypes
                             {[\LIT{data}\ \tau = [c\ \tau_i]]}
                             {[c\ p_i]}
                             {\tau}
                             {(\circ \Sigma_i)}
                           }
\end{center}
\caption{Typing environment inference.}
\label{fig:variable-type-bindings}
\end{figure}

\begin{figure}
    \begin{flushleft}
      \fbox{\LinearUnBindTypes{}{t}{\tau}{\Sigma}}
    \end{flushleft}
\begin{center}
  \Axiom
  {\LinUNFER{Variable}}    
                           {\LinearUnBindTypes{}{x}{\tau}{\{x \mapsto \tau\}}}

  \InfOne                  {\LinearUnBindTypes{}{t_i}{\tau_i}{\Sigma_i}}
  {\LinUNFER{Constructor}} 
                           {\LinearUnBindTypes
                             {[\LIT{data}\ \tau = [c\ \tau_i]]}
                             {[c\ t_i]}
                             {\tau}
                             {(\circ \Sigma_i)}
                           }

  \InfThree                {\LinearUnBindTypes{}{t_i}{\tau}{\Sigma_{t_i}}}
                           {\LinearBindTypes{}{p_i}{\tau_p}{\Sigma_{p_i}}}
                           {\LinearUnBindTypes{}{t}{\tau_p}{\Sigma}}
  {\LinUNFER{Cases}}       
                           {\LinearUnBindTypes
                             {}
                             {\CASE{t}{\tau_p}{p_i}{t_i}}
                             {\tau}{(\Sigma_{t_i} - \Sigma_{p_i}) \circ \Sigma}
                           }
                           $(\psi)$

  \InfOne                  {\LinearUnBindTypes{}{p}{\tau'}{\Sigma}}
  {\LinUNFER{Application}} 
                           {\LinearUnBindTypes
                             {[f (\cdot : \tau) : \tau' = \cdot]}
                             {f p}
                             {\tau}
                             {\Sigma}}

  \InfOne                  {\LinearBindTypes{}{\APP{g}{p}}{\tau}{\Sigma}}
  {\LinUNFER{Inversion}}   
                           {\LinearUnBindTypes{}{\APP{\INVERT{g}}{p}}{\tau}{\Sigma}}
\end{center}
\caption{Inverse typing environment inference.}
\label{fig:inverse-variable-type-bindings}
\end{figure}

As mentioned in Section~\ref{sec:language-description}, running a program
corresponds to applying its main function to a value provided by the caller
in an empty context. So, the desirable property for programs to have, is
that this application yields a unique result, and that calling the inverted
program on the result will yield said provided input. This property has been
summarised in Theorems~\ref{thm:backwards-determinism} and~\ref{thm:reversible-semantics}, and the nifty Corollary~\ref{thm:locally-invertible}.

\begin{theorem}
  \label{thm:backwards-determinism}
  If $t$ is a linear term then \EvalWith{}{\Gamma}{t}{v} if and only if
  \Because{}{t}{v}{\Gamma}.
\end{theorem}
\renewcommand*{\proofname}{Proof outline}
\begin{proof}
  By induction on the derivations $\mathcal D$ of \EvalWith{}{\Gamma}{t}{v}
  and $\mathcal C$ of \Because{}{t}{v}{\Gamma} respectively. Here we give
  the case for function application:
  \begin{itemize}
  \item Suppose $t$ is a function applied in the conventional
    direction. Then $t$ looks like $f~p$, and $\mathcal D$ is a derivation
    of \EvalWith{}{\Gamma}{f~p}{v}, and so it must have used the
    \DOWNARROW{Application} rule. As such, $\mathcal D$ must be constructed
    from a derivation $\mathcal D_1$ of \EvalWith{}{\Gamma}{p}{v'} and
    another derivation $\mathcal D_2$ of \EvalWith{}{(\MGU{v'}{p'})}{t'}{v}
    where $p'$ and $t'$ are the argument pattern and function body of $f$ as
    defined in $\Delta$.  Furthermore, by the definition of \MGU{v'}{p'} and
    the fact that $v'$ is a value (and thereby variable-free): If $p'$ is a
    variable, then \EvalWith{}{(\MGU{v'}{p'})}{p'}{v'} holds by the
    \DOWNARROW{Variable} rule, and otherwise, $p'$ is a constructor and we
    can use the \DOWNARROW{Constructor} rule to obtain the same proof. In
    either case, we can construct a derivation $\mathcal D_3$ of
    \EvalWith{}{(\MGU{v'}{p'})}{p'}{v'}.

    Now, by the induction hypothesis on $\mathcal D_1$, we get a derivation
    $\mathcal C_1$ of \Because{}{p}{v'}{\Gamma}, and by the induction
    hypothesis on $\mathcal D_2$, we get a derivation $\mathcal C_2$ of
    \Because{}{t'}{v}{\MGU{v'}{p'}}.  And finally, we can apply the
    \RIGHTARROW{Application} rule to $\mathcal C_2$, $\mathcal D_3$ and
    $\mathcal C_1$ respectively we obtain a derivation $\mathcal C$ of
    \Because{}{f~p}{v}{\Gamma}.

    Conversely, we can throw away $\mathcal D_3$ and use the induction
    hypothesis on $\mathcal C_1$ and $\mathcal C_2$ to reconstruct $\mathcal
    D_1$ and $\mathcal D_2$, which we may use reconstruct $\mathcal D$ and we are done.
  \end{itemize}
    The difficult bit is to show uniqueness of evaluation for
    \DOWNARROW{Cases}, by unfolding the meaning of the bi-directional first
    match policy $\psi$ specified in bullet point (\ref{rule:eval-case}).
\end{proof}

\begin{theorem}
  \label{thm:reversible-semantics}
  Let $\Delta[f (p : \tau_p) : \tau_t) = t.]$ be a program in which a
  function $f$ has been declared, and consider two values $v$ and $w$ such
  that \LinearHasType{}{\emptyset}{v}{\tau_p} and
  \LinearHasType{}{\emptyset}{w}{\tau_t} holds. Then
  \EvalWith{}{\emptyset}{\APP{f}{v}}{w} if and only if
  \LaveWith{}{\emptyset}{\APP{f}{w}}{v}.
\end{theorem}
\renewcommand*{\proofname}{Proof}
\begin{proof}
  Suppose \EvalWith{}{\emptyset}{\APP{f}{v}}{w}, then the derivation must
  have used the \DOWNARROW{Application} rule. Consequently, we get a
  derivation derivation of \EvalWith{}{\MGU{v}{p}}{t}{w}. Now, by Theorem
  \ref{thm:backwards-determinism}, we get another derivation of
  \Because{}{t}{w}{\MGU{v}{p}}, and from the definition of the most general
  unifier, we derive \EvalWith{}{\MGU{v}{p}}{p}{v}. Since $w$ is a value,
  clearly \EvalWith{}{\emptyset}{w}{w}. So, we can apply the
  \UPARROW{Application} rule to show that
  \LaveWith{}{\emptyset}{\APP{f}{w}}{v}. The converse proof is similar.
\end{proof}

\noindent
From this theorem and the \DOWNARROW{Invert} rule follows that inversion
is well-behaved:

\begin{corollary}
  \label{thm:locally-invertible}
  Let $\Delta$ be a program, and $f$ an arbitrary function defined in
  $\Delta$. Then \Eval{\APP{f}{v}}{w} if and only if \Eval{\APP{\INVERT{f}}{w}}{v}.
\end{corollary}

\section{Implementing Invertible Semantics}
\label{sec:invertible-semantics}

So far we have investigated several competing ideas for implementing
the invertible semantics of Jeopardy. We have seen in
Section~\ref{sec:examples} that in some cases, it is sufficient to
extend the bi-directional first match policy of RFun and CoreFun to
include information that is implicitly provided by the caller. We have
developed a program analysis, based on the \textit{available
  expressions analysis} specified in Nielsen, Nelson, and
Hankin~\cite{NielsonFlemming2015PoPA}, called \textit{implicitly
  available arguments analysis} which was presented at NIK
2022~\cite{KristensenKaarsgaardThomsenNIKPaper}. Based on this
analysis, one can extend the judgements for evaluating programs with a
statically available environment that contains bindings provided by
the caller, and the side condition in Figures~\ref{fig:evaluation}
and~\ref{fig:inference} will be an extended notion of orthogonality
that is allowed to look at the pattern in each case as
well. Additionally, available implicit arguments analysis leads the
way for a number of program transformations that compile away certain
branching constructions in which branching symmetry is not locally
decidable (e.g. by syntactic orthogonality). The information provided
by this particular analysis can also be used to prune the search space
of algorithms, such as the one found in Figure~\ref{fig:inference}; we
also conjecture that online partial evaluation can be used to
eliminate branches from a case-statement that do not agree with the
implicitly provided arguments in the opposite direction of
interpretation. An extended type system could also use available
expressions to generate a set of constraints in order to
conservatively verify that the program complies with this extended
notion of term-pattern orthogonality, and that $\psi$ does not need to
be checked at runtime in such cases.

The caveat of this kind of analysis is that it is syntax directed, and
as seen with the \texttt{map-f} example in Section~\ref{sec:examples},
this is not always sufficient. Several strategies are available to us
to extend this:
\begin{itemize}
  \item We can generalise the program transformation employed for
    \texttt{map-f-iter} in Section~\ref{sec:examples}, that generalises the
    case where the caller provides information about something that the
    callee does not explicitly branch over.
  \item We allow free \textit{existential} variables (unification variables)
    that we know we can find by unification some time in the future (before
    it is needed). That is, functions behave like Horn clauses which are
    known (by a separate program analysis) to succeed exactly once, that
    guarantees all existential variables are bound (i.e., constrained to be
    equal to a ground term) by the time they return.
  \item Instead of a syntax directed analysis, we can instead produce
    a graph structure in the style of
    \cite{LeeJonesBenAmram:2001}. Using such an analysis, we can
    enforce that invertible functions terminate in a unique case for
    each possible constructor in the data type definition for its
    input. One could even construct a list of graphs from elaborated
    Jeopardy programs that give names to the parts of the program that
    are not bound. These graphs could then be used as a more expressive
    alternative to syntax in further analyses and transformations.
\end{itemize}

\section{Conclusion}
\label{sec:conclusion}
The study of invertible computation has, historically, proven useful in
understanding energy and entropy preservation, and in understanding
information preserving transformations and transmission. However, there is
still something to be learned about programs that are invertible, in particular
regarding how to make invertible programming less syntactically restrictive,
and how to implement this in a reasonably efficient way.

Since program invertibility is undecidable in
general, all an invertibility analysis can hope to achieve is a reasonable
approximation. In other words, any static  analysis will
split the expressible programs into three groups: those which are found to be
invertible, those which are found to \emph{not} be invertible, and those for
which the analysis can provide no definite answer.

Clearly, the goal of our work is to make this latter class of programs as
small as possible. However, since RFun and CoreFun are both R-Turing
complete languages, we cannot hope to achieve a more computationally
powerful language, though we \emph{can} hope to make an invertible language
that is more concise, familiar, and user friendly by enabling the expression
of algorithms in a style that is much closer to that of conventional functional
programming languages.

\bibliographystyle{abbrv}
\bibliography{sample-base}

\end{document}